\documentclass[twocolumn,superscriptaddress,amsmath,amssymb,aps,prb,longbibliography]{revtex4-2}
\usepackage[utf8]{inputenc}
\usepackage{amssymb}
\usepackage{amsmath}
\usepackage{braket}
\usepackage[english]{babel}
\usepackage{graphicx}

\usepackage{tikz}
\usepackage{graphicx,bm}
\usetikzlibrary{positioning}

\usepackage{indentfirst}
\date{May 2022}

\begin{document}
\title{Magnetotransport on quantum spin Hall edge coupled to bulk midgap states}
\author{Youjian Chen}
\affiliation{Department of Physics, University of Virginia, Charlottesville, VA 22904, USA}
\author{Wenjin Zhao}
\affiliation{Kavli Institute at Cornell for Nanoscale Science, Ithaca, NY, USA}
\author{Elliott Runburg}
\affiliation{Department of Physics, University of Washington, Seattle, USA}
\author{David Cobden}
\affiliation{Department of Physics, University of Washington, Seattle, USA}
\author{D. A. Pesin}
\affiliation{Department of Physics, University of Virginia, Charlottesville, VA 22904, USA}
\begin{abstract}
    We consider magnetotransport on a helical edge of a quantum spin Hall insulator, in the presence of bulk midgap states ``side-coupled" to the edge. In the presence of a magnetic field, the midgap levels are spin-split, and hybridization of these levels with the itinerant edge states leads to backscattering, and the ensuing increase in the resistance. We show that there is a singular cusp-like contribution to the positive magnetoresistance stemming from resonant midgap states weakly coupled to the edge. The singular behavior persists for both coherent and incoherent edge transport regimes. We use the developed theory to fit the experimental data for the magnetoresistance for monolayer WTe$_2$ at liquid helium temperatures. The results of the fitting suggest that the cusp-like behavior of the resistance in weak magnetic fields observed in experiments on monolayer WTe$_2$ with long edge channels might indeed be explained by hybridization of the helical edge states with spin-split bulk midgap states. In particular, the dependence of the magnetoresistance on the direction of the external magnetic field is well described by the incoherent edge transport theory, at the same time being quite distinct from the one expected for a magnetic-field-induced edge gap. 
\end{abstract}
\maketitle

\section{Introduction}
Transport and magnetotransport on a helical edge of a quantum spin Hall, or two-dimensional topological insulator, phase have been considered one of the main tests for the identification of this phase since its theoretical proposal\cite{kane2005quantum,kane2005z,bernevig2006quantum} and initial experimental discoveries~\cite{konig2007quantum}. 
 
Traditionally, considerations of the edge conductance focus on its two related aspects: the conductance quantization (or often lack thereof) at the value of $2e^2/h$ at low temperatures, and its suppression by various perturbations, magnetic fields in particular. Arguably, the conductance quantization issue has received more theoretical and experimental attention. If one accepts the view that ideally the conductance on a helical edge must be quantized at sufficiently low temperatures, the main issue to address is the observed lack of quantization. There are several proposed mechanisms for this: localization of electrons by magnetic impurities~\cite{zhang2009kondo,matveev2011kondo,aleiner2013localization}, vacancy-induced magnetic moments~\cite{polini2019magnetic}, the existence of charge puddles~\cite{vayrynen2013puddles,essert2015magnetotransport}, interaction-mediated backscattering~\cite{moore2006interaction,glazman2012backscattering}, spontaneous TR-breaking on the edge~\cite{gefen2017trbreaking}, backscattering by nuclear magnetic moments~\cite{loss2017nuclear}, or noise~\cite{alicea2018noise}.

In this paper we completely put aside the issue of the conductance quantization and focus solely on its suppression by an external magnetic field. Given the above list of mechanisms proposed to explain the lack of conductance quantization, it is likely that the number of those affecting the magnetoconductance is just as extensive. Therefore, it makes sense to limit the scope of our study to the minimal set of physical aspects of the problem that can explain the observed features of the edge transport. We thus limit ourselves to considerations of elastic transport (coherent or incoherent) in the presence of coupling between the edge electrons and either potential disorder on the edge or resonant mid-gap levels in the bulk.

We specifically focus on explaining the observed magnetoconductance in $1\rm{T}'$-WTe$_2$ monolayers. (Without a magnetic field, but in the presence of a strong disorder, the edge conductance of this material was theoretically discussed in Ref.~\cite{bieniek2022theory}.) Monolayer WTe$_2$ was theoretically proposed to host the quantum spin Hall insulator phase in Refs.~\cite{Fu_2014,Chang_2016,Xiang_2016,Zheng_2016}. Subsequent experimental studies confirmed the presence of edge conduction at low temperatures~\cite{Cobden_2017,Tang_2017,Jia_2017,Peng_2017,wu2018wte2,Song_2018,Shi_2019,Pesin_2020,hamilton2021spin}. 

Detailed edge magnetoconductance studies~\cite{Pesin_2020} showed that the edge conductance strongly depends on the magnetic field orientation with respect to a well-defined direction, which essentially does not depend on sample, gate voltage, or edge orientation in a given sample. It is natural to identify this special direction, canted away from the normal to the sample due to the low symmetry of the latter, with the direction of the spin polarization of the spin-momentum locked edge states~\cite{Pesin_2020,hamilton2021spin,garcia2020canted,nandy2022kp}. 

The edge magnetoresistance in WTe$_2$ for small B-fields appears to have a non-analytic cusp-like field dependence, somewhat rounded, as it must be, at the smallest B-fields. Furthermore, the conductance is suppressed by roughly an order of magnitude in magnetic fields of order of 0.1-0.5 Tesla, for field orientations perpendicular to the edge spin polarization. The strong field dependence of the edge conductance at small B-fields is puzzling. 

There are various ways to explain such strong magnetic field dependence. One can broadly separate the effects associated with a magnetic field into spin (``Zeeman") and orbital ones. The orbital mechanisms of conductance suppression are strongest for magnetic fields oriented perpendicular to the two-dimensional sample plane. They are thought to be behind the strong conductance suppression by out-of-plane magnetic fields in HgTe/CdTe quantum wells~\cite{konig2007quantum}. Naturally, in order to have an appreciable effect on transport, weak orbital magnetic field requires a way to produce a large magnetic flux, either through a charge puddle near the edge, or through some effective loop formed by an edge state forced to ``reptate" away from the physical edge by strong disorder ~\cite{zhang2010numerics,buttiker2012localization,pikulin2014,essert2015magnetotransport}.

Unlike the case of HgTe/CdTe quantum wells, there is no strong qualitative difference between edge magnetoconductance in WTe$_2$ for in-plane and out-of-plane magnetic fields. Due to the canted spin polarization on the edge, both of the above field orientations make substantial angles with the edge spin polarization (roughly $55^\circ$ and $35^\circ$, respectively), and show similar magnetoconductance.  This appears to be consistent with strong lateral confinement of the edge states in WTe$_2$ due to a large direct gap~\cite{Neupert2016prx,Lau_2019,thomale2019prb,thomale2022edge}, as distinct from the actual narrow indirect gap of the material.

If Zeeman fields are responsible for edge magnetoconductance in WTe$_2$, then the most obvious mechanism for conductance suppression would be opening up a gap at a topological edge. If the chemical potential happens to be very close to the Dirac point energy on the edge, magnetic field will indeed lead to insulating behavior, as conjectured in Ref.~\cite{wu2018wte2} to explain observed magnetoconductance for very short channels. However, the cusp-like behavior of the mangetoconductance persists for a wide range of gate voltages~\cite{Pesin_2020,hamilton2021spin}, which suggests that the gap opening on the edge is at the very least not the only mechanism of singular magnetoconductance. 

Converted to energy units, magnetic fields of $0.1-0.5$ Tesla correspond to an energy scale of 0.1 meV (or 1 K in temperature units), unless the $g$-factor on the edge is anomalously large. There are no reports of such edge $g$-factors to the best of our knowledge. It is then a key question how such small fields can have a large effect on the edge conductance for generic values of the chemical potential, say, from meVs to tens of meVs away from an edge Dirac point.  It is hard to find a relevant energy scale that controls edge conduction and which is small enough to be overcome by such a small magnetic field. 

In this paper we argue that small Zeeman fields can be given a chance to have a large effect on transport if they are able to cause spin precession that persists for a long time. It is impossible to achieve this on the edge itself, because of the strong spin-momentum locking and the associated strong effective Zeeman fields coming from the spin splitting at the Fermi level. However, edge electrons can be hybridized with bulk mid-gap states in a way roughly described by the (noninteracting) Anderson impurity model. In WTe$_2$, such states can come from Te-vacancies~\cite{muechler2020impurities}. Then edge electrons can spend considerable time in the bulk states while traversing an edge. Since mid-gap states are Kramers-degenerate for zero fields, (we assume that the electron-electron interaction is weak, and spin magnetic moments are not formed), then even small Zeeman fields will cause large precession angles for electrons spending time on resonant bulk mid-gap states. In this paper we show that this can lead to significant backscattering, and conductance suppression with a cusp-like behavior for small fields. 

\begin{figure}
    \centering
    \includegraphics[width=3.5in]{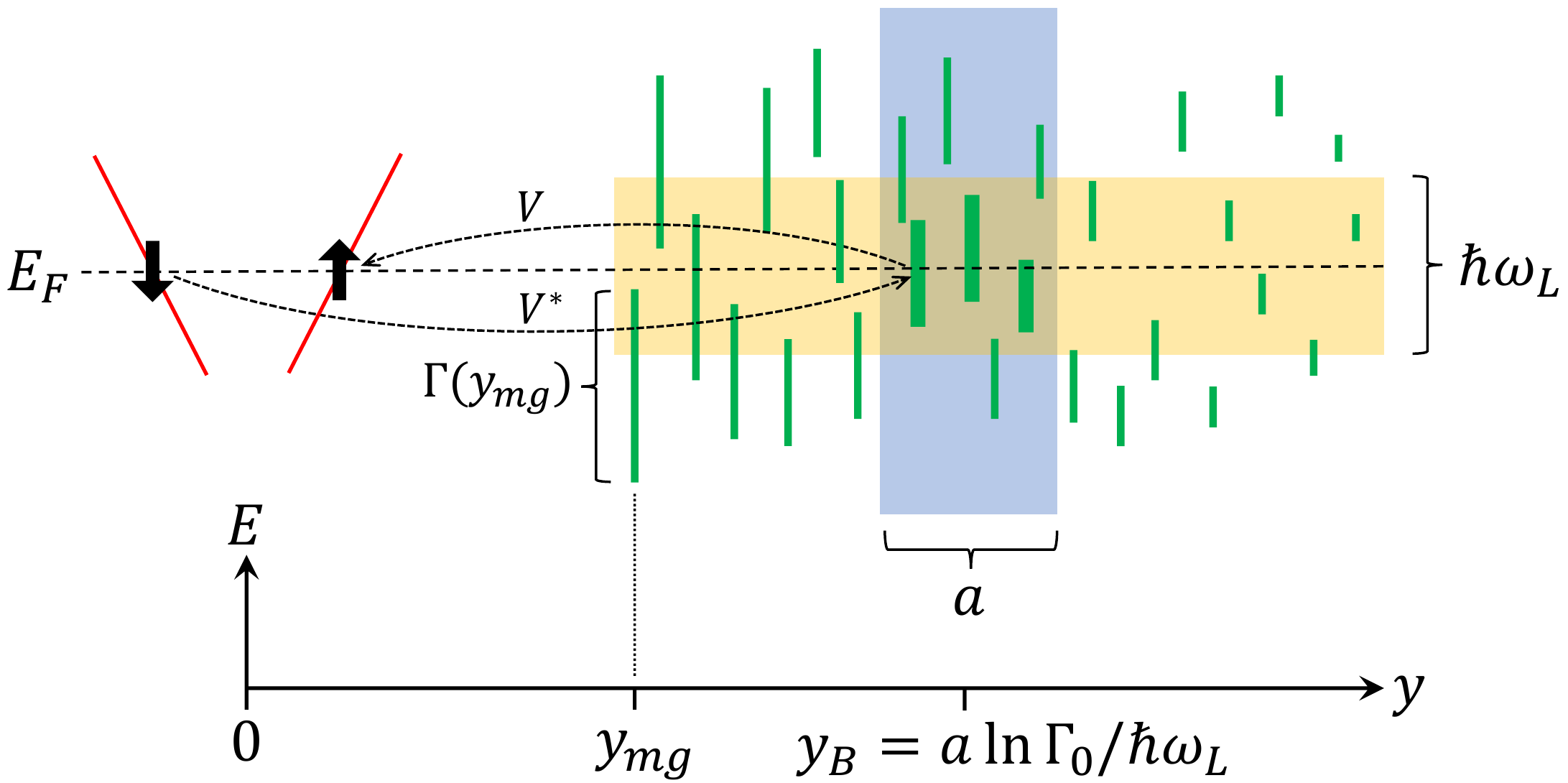}
    \caption{Schematic view of helical edge states hybridized with bulk midgap states. The vertical axis is energy and the horizontal axis, $y$, is distance from the physical edge. Each vertical bar represents a midgap state: its $y$-coordinate,  $y_{mg}$, describes its localization center in real space; its middle is the energy of the level; and its length $\Gamma(y_{mg})$ indicates the width of the level due to hybridization with the edge states (the matrix elements of which are $V$ and $V^*$). The vertical blue shaded band indicates the range in which $\Gamma(y_{mg})$ is comparable to $\hbar\omega_L$, where $\omega_L$ is the Larmor precession frequency for a typical midgap state. In real space, these states are located in a strip centered at $y_B$ whose width $a$ is roughly the typical extent of a midgap state wave function. The horizontal (yellow) shaded band indicates the levels that are within $\hbar\omega_L$ of the Fermi level, $E_F$. The states (bold green bars) within the intersection of the two bands backscatter edge electrons with a probability of order unity. The number of such states is proportional to $\hbar\omega_L$, leading to a distinctive $|B|$ behavior of the magnetoresistance at small fields.} 
    \label{fig:edge-bulk}
\end{figure}

 A simple qualitative picture of the strong edge conductance suppression by a magnetic field in the presence of midgap bulk levels can be derived from a semiclassical argument based on the field-induced spin precession of electrons residing on a midgap level due to its hybridization with the edge state. A cartoon of helical edge states coupled to bulk midgap states, which are distributed in both real space and energy, is shown in Fig.~\ref{fig:edge-bulk}.

 First, consider a midgap level resonant with a scattering electron traversing a helical edge of a quantum spin Hall insulator. Without a magnetic field, even for nonzero hybridization of a midgap level with the edge states, it can only lead to forward scattering of edge electrons due to the helical nature of the edge state, and time-reversal symmetry\cite{maciejko2011review}. However, a scattering edge electron wave packet spends the Wigner delay time on the midgap level before going back into an outgoing edge mode. (This picture also holds in the presence of a magnetic field.) If the width of the level due to its hybridization with the edge electrons is denoted as $\Gamma$, then the time an electron spends on a resonant impurity level  is $~\sim \hbar/\Gamma$. For a small magnetic field with a finite component perpendicular to the edge spin polarization, one can think of electron spin precession while it resides on the midgap level, hence the spin orientation of the electron rotates by an angle $\sim \hbar\omega_L/\Gamma$, where $\omega_L$ is the Larmor precession frequency for the midgap level, in the sense that it involves the g-factor for the level. If the spin of an electron rotates by an angle of order unity from its original orientation, then the electron can be re-scattered into both helical channels. This implies that midgap levels whose energies are within a window of $\hbar \omega_L$ of the Fermi level, and whose widths $\Gamma\sim\hbar \omega_L$ backscatter electrons with a probability close to one, leading to substantial conductance suppression. Midgap levels that are well outside this window, or whose width is too different from the spin splitting, are essentially decoupled from the edge states at the Fermi level. 
 
 A real sample will have many midgap levels, see Fig.~\ref{fig:edge-bulk}. The level width due to coupling to the edge states varies with the distance to the physical edge of the system. Formally, for any small magnetic field there is a group of midgap levels far enough from the edge that their width is comparable to the spin splitting energy due to the magnetic field. If these levels are also close to the Fermi level, they suppress the edge conductance efficiently. It is intuitively clear that the strength of conductance suppression is determined by the number of such strongly backscattering midgap states. This number can be estimated by counting the number of midgap levels within an energy window of $~\hbar \omega_L$ around the Fermi level, and in real space within a strip of width determined by the condition that $\Gamma$ is not too different from $\hbar \omega_L$ (not too small, and not too large). Let us denote the level width as a function of the distance to the edge, $y$, as $\Gamma(y)$. Then the distance of this real space strip from the edge, $y_B$, is found from $\Gamma(y_B)\sim \hbar \omega_L$, and its width is $(dy_B(\Gamma)/d\Gamma)\hbar\omega_L$. We use the simplest, but physically motivated, possible model of level width, $\Gamma(y)=\Gamma_0\exp(-y/a)$, with $\Gamma_0$ determining the maximum width of the level in the vicinity of the physical edge and $a$ the scale on which the level width decays in the bulk. We obtain $y_B=a\ln(\Gamma_0/\hbar\omega_L)$, while the strip width is field-independent and is simply $a$, such that $y_B\gtrsim a$ always holds at small magnetic fields. Then, under the plausible assumption that the density of midgap levels is approximately uniform near the Fermi level, we see immediately that the number of resonant levels goes linearly with $\hbar \omega_L$, and hence linearly with the magnetic field magnitude. This is the essential origin of the striking cusp displayed by the magnetoconductance at zero field. 

 Within the above picture, the scale of the rounding of the cusp is determined as follows. If the field is so weak that the resonant midgap levels are very far from the edge, and the required level width is smaller than the one due to inelastic processes like phonon-assisted hopping, then we expect the preceding considerations not to apply. It is hard to estimate the corresponding field scale, but it can be read off from the data. 

The rest of the paper is organized as follows. We consider backscattering on a helical edge in the presence of a magnetic field
due to a single bulk mid-gap state in Section~\ref{sec:impscattering}. In Section~\ref{sec:transport} we calculate the magnetoconductance of a helical edge  in coherent and incoherent regimes, and show that the presence of resonant mid-gap states leads to cusp-like feature in the conductance at small B-fields. Finally, we discuss our results and their relation to the experimental data on magnetotransport in monolayer WTe$_2$ in Section~\ref{sec:discussion}.

\section{Disorder scattering on a helical edge}\label{sec:impscattering} 

In this section, we consider helical edge electrons coupled to a midgap state localized somewhere in the bulk, but hybridized with the edge states. If the energy of the midgap state is sufficiently close to the Fermi level, electrons spend a considerable time in that state while traveling along the edge. In the presence of a magnetic field misaligned with the edge spin polarization, the spin of the electron on the midgap level precesses, and allows backscattering on the edge.  The characteristic field in this case is such that the corresponding Zeeman energy is comparable to the level width due to coupling to the edge. For impurities not too close to the edge, the width can be very small, in particular much smaller than the spin splitting at the Fermi level, hence small magnetic fields can induce substantial backscattering. 

In Appendix~\ref{sec:potentialscattering} we will contrast the above scenario with a more conventional model of a short-range potential impurity on the edge. Without magnetic field, there is no backscattering due to the TR-symmery of the problem. Nonzero magnetic field acts to tilt the spins of the helical states in the same direction, leading to a finite overlap, and allowing backscattering. The important observation from this example is that the Zeeman energy associated with the B-field must be comparable to the zero-field spin splitting at the Fermi level in order to lead to significant backscattering. As a result, this type of scattering cannot lead to non-analytic magnetoconductance behavior in small fields.

The case of conventional wires ``side-coupled" to an impurity level was considered in Ref.~\cite{lerner2008impurity}. Here we consider this problem for a helical edge state coupled to a midgap electronic state located in the bulk of the quantum spin Hall insulator. 

The Hamiltonian of the problem has three parts:
\begin{equation}\label{eq:totalH}
    H_{tot}=H_{edge}+H_{hyb}+H_{mg}.
\end{equation}
Here $H_{edge}$ is the Hamiltonian of edge electrons with momentum $k$ along the edge direction, Dirac speed $v$, and spin projections $\sigma=\uparrow,\downarrow$ on a particular axis that defines the edge spin polarization at the chemical potential $\mu$. While the problem we are solving is single-particle, it is still convenient to express the corresponding Hamiltonians in the second-quantized form. We then have
\begin{align}
    H_{edge}=\sum_{k\sigma}(\sigma v k-\mu)a^\dagger_{k,\sigma}a_{k,\sigma},
\end{align}
where we treat $\sigma=\uparrow,\downarrow$ as $\pm1$, respectively, when used in an equation, $a_{k,\sigma}$ is the annihilation operator for the electrons with momentum $k$ and spin projection $\sigma$. In this Section we ignore the Zeeman coupling for the edge electrons, assuming the corresponding energy scale is much smaller than the spin splitting of the edge states at the Fermi level. Inclusion of the Zeeman coupling will only yield unimportant perturbative effects, quadratic in the magnetic field.  

To describe the edge-midgap state hybridization, we will assume that the midgap state is localized around $x_{mg}=0$, $x$ being the coordinate along the edge. This allows us to avoid factors like $\exp(ikx_{mg})$ in the hybridization matrix element. With this assumption, $H_{hyb}$ is given by 
\begin{equation}
    H_{hyb}= \sum_{k,\sigma}( V a^{\dagger}_{k,\sigma} c_{\sigma}+V^{*}c_{\sigma}^{\dagger} a_{k,\sigma}),
\end{equation}
where $c_{\sigma}$ is the annihilation operator for electron with spin $\sigma$ in the midgap state, and the hybridization matrix element $V$ is assumed to be spin conserving and momentum independent. 

Finally, $H_{mg}$ describes a midgap level with energy $\epsilon_{mg}$, and includes Zeeman coupling to a magnetic field $\bm {B}$. We will use the notation $\bm{b}=\frac12 g_{mg}\mu_{B}\bm{B}\equiv (b_{x},b_{y},b_{z})\equiv(b \sin \theta \cos \phi, b \sin \theta \sin \phi, b \cos \theta)$, where $g_{mg}$ is the $g$-factor for the midgap state. For definiteness, we measure angle $\theta$ of a spherical coordinate system from the ``up" direction of the edge spin. With this notations, $H_{mg}$ is written as
\begin{equation}
    H_{mg}=\begin{pmatrix}
     c^{\dagger}_{\uparrow} &
    c^{\dagger}_{\downarrow} \\
    \end{pmatrix} 
    \begin{pmatrix}
    \epsilon_{d}+b_{z} & b_{x}-i b_{y} \\
    b_{x}+i b_{y} & \epsilon_{d}-b_{z} \\
    \end{pmatrix}
    \begin{pmatrix}
    c_{\uparrow} \\
    c_{\downarrow}
    \end{pmatrix}.
\end{equation}

We now proceed to solve the scattering problem for Hamiltonian~\eqref{eq:totalH}, treating $H_{hyb}$ as the scattering part. It is convenient to describe the midgap state using the eigenstates of $H_{mg}$, modifying the hybridization accordingly.

The single-particle Hamiltonian of the midgap state can be diagonalized as 
\begin{equation}
    U  \begin{pmatrix}
    \epsilon_{d}+b_{z} & b_{x}-i b_{y} \\
    b_{x}+i b_{y} & \epsilon_{d}-b_{z} \\
    \end{pmatrix}
    U^{\dagger}=
    \begin{pmatrix}
\epsilon_{+} & 0\\
0 &\epsilon_{-} \\
    \end{pmatrix}
\end{equation}
where the eigenvalues are $\epsilon_{\pm}=\epsilon_{d}\pm b$. The unitary matrix $U$ diagonalizing the midgap state Hamiltonian can be defined as connecting the annihilation operators for the spin states of the midgap level, $c_{\uparrow,\downarrow}$, to those of the eigenstates of the midgap level Hamiltonian, $c_\pm$:
\begin{equation}
    \begin{pmatrix}
    c_{\uparrow} \\
    c_{\downarrow}
    \end{pmatrix}
    =
    U^{\dagger}
    \begin{pmatrix}
    c_{+} \\
    c_{-}
    \end{pmatrix}
\equiv
\begin{pmatrix}
\cos \frac{\theta}{2} & -\sin \frac{\theta}{2} e^{-i \phi }\\
    \sin \frac{\theta}{2} e^{+i \phi} &\cos\frac{\theta}{2} \\
\end{pmatrix}
\begin{pmatrix}
    c_{+} \\
    c_{-}
    \end{pmatrix}.
\end{equation}
In the new basis, the midgap level Hamiltonian is written as 
\begin{equation}
    H_{mg}=
    \begin{pmatrix}
    c^{\dagger}_{+} &
    c^{\dagger}_{-}
    \end{pmatrix}
    \begin{pmatrix}
\epsilon_{+} & 0\\
0 &\epsilon_{-} \\
    \end{pmatrix}
    \begin{pmatrix}
    c_{+} \\
    c_{-}
    \end{pmatrix},
\end{equation}
while the hybridization between the edge electrons with spins $\uparrow,\downarrow$, and the midgap level eigenstates labeled with $\pm$ becomes
\begin{equation}
    V_{\sigma s}=
    \begin{pmatrix}
        V_{\uparrow+} & V_{\uparrow-}\\
         V_{\downarrow+} & V_{\downarrow-} \\
    \end{pmatrix}
    =
   V \begin{pmatrix}
        \cos\frac{\theta}{2} & -\sin\frac{\theta}{2}e^{-i\phi} \\
        \sin\frac{\theta}{2}e^{+i\phi} & \cos\frac{\theta}{2}\\
    \end{pmatrix},
\end{equation}
such that the hybridization Hamiltonian is
    \begin{equation}
    H_{hyb}=\sum_{k,\sigma=\uparrow,\downarrow}\sum_{s=+,-}V_{\sigma s} a_{k, \sigma}^{\dagger}c_{s}+h.c.
\end{equation}

Since we are dealing with a single-particle problem, in what follows we switch to single-particle retarded Green's function, and the T-matrix to solve the corresponding Schr{\"o}dinger equation using the Lippmann-Schwinger approach. 

We first define a few quantities for the case of uncoupled edge and midgap level. The single-particle retarded Green's function can be written as the sum of single-particle operators acting in the the edge and midgap level subspaces of the full electronic Hilbert space that contains the edge and midgap states: 
\begin{equation}
    \hat{G}_{0}=\hat{G}_{edge}+\hat{G}_{mg}
\end{equation}
The edge Green's function is written as
\begin{equation}\label{eq:edgeGF}
    \hat{G}_{edge}=\sum_{k\sigma=\uparrow,\downarrow}\frac{\ket{k ,\sigma} \bra{k ,\sigma}}{E+i \eta - \sigma v k},
\end{equation}
where $\eta$ is a positive infinitesimal, and $\ket{k\sigma}$ are the single-particle eigenstates of $H_{edge}$ with momentum $k$ and spin $\sigma$. The midgap level Green's function is
\begin{equation}
    \hat{G}_{mg}=\sum_{s=\pm}\frac{\ket{s}\bra{s}}{E+ i\eta-\epsilon_{s}},
\end{equation}
where $\ket{s}$  with $s=\pm$ are the midgap states with energies $\epsilon_\pm$.

The hybridization between the edge and midgap states can be decomposed into a sum of two parts, $\hat{V}=\hat{\mathcal{V}}+\hat{\mathcal{V}}^{\dagger}$. The operator $\mathcal{V}$ only has matrix elements for transitions from the midgap state into edge states:
\begin{equation}
\hat{\mathcal{V}}=\sum_{k}\sum_{s,\sigma}V_{\sigma s}\ket{k,\sigma}\bra{s},
\end{equation}
while $\hat{\mathcal{V}}^{\dagger}$ only has matrix elements for transitions from the edge states to the midgap level:
\begin{equation}
\hat{\mathcal{V}}^{\dagger}=\sum_{k} \sum_{s,\sigma}V_{\sigma s}^{*} \ket{s}\bra{k,\sigma}.
\end{equation}
From now on we will suppress the `hats' on top of single-particle operators. 

The equation for the T-matrix in the full electronic Hilbert space, $T_{full}$, has the usual form of   $T_{full}=\sum_{n=0}^{\infty}\hat{V}(\hat{G}_{0}\hat{V})^{n}$. By solving for the matrix elements of the T-matrix that couple the edge and midgap states, one can obtain the T-matrix for the edge electrons, $T$:
\begin{equation}
    T=\hat{{\mathcal{V}}}\hat{G}_{mg} \hat{\mathcal{V}}^{\dagger} (1-\hat{G}_{edge}\hat{\mathcal{V}}\hat{G}_{mg}\hat{\mathcal{V}}^{\dagger})^{-1}.
\end{equation}
\begin{widetext}

Using the definition $T^{ k_ak_b}_{ \sigma_{a}\sigma_{b}}=\bra{k_{a}\sigma_{a}} T \ket{k_{b}\sigma_{b}}$ to describe the scattering matrix elements between edge states $\ket{k_{a}\sigma_{a}}$ and $\ket{k_{b}\sigma_{b}}$, we can get an algebraic equation for the edge T-matrix from the Lippmann-Schwinger equation:
\begin{equation}\label{eq:Tmatrixeq}
T^{k_{a}k_{b}}_{\sigma_{a}\sigma_{b}}-\sum_{k}\sum_{s,\sigma } \frac{T^{k_{a}k_{b}}_{\sigma_{a}\sigma}V_{\sigma s}V_{\sigma_{b} s}^{*}}{(E+i \eta-\sigma v k)(E+i\eta-\epsilon_{s})}=\sum_{s}\frac{V_{\sigma_{a} s}V_{\sigma_{b} s}^{*}}{E+i \eta-\epsilon_{s}}.
\end{equation}
Since the hybridization matrix elements are momentum independent, the T-matrix should also be momentum independent $T^{k_{a}k_{b}}_{\sigma_{a}\sigma_{b}}=T_{\sigma_{a}\sigma_{b}}$. 
Then the sum over momenta $k$ in Eq.~\eqref{eq:Tmatrixeq} can be performed by switching to integration: $\sum_{k}\to\frac{L_{x}}{2 \pi} \int dk$. According to the Sokhotski–Plemelj theorem, $\frac{1}{E+i\eta-\epsilon_{k \sigma}}=P\frac{1}{E-\epsilon_{k\sigma}}-i\pi \delta(E-\epsilon_{k\sigma})$. The principal integral has the meaning of the energy shift of the midgap level due to the coupling to the edge electrons. We will assume that this shift has already been incorporated into $\epsilon_s$, and disregard the principal value integral. We then obtain the final equation for the edge T-matrix:
\begin{equation}
    T_{\sigma_{a}\sigma_{b}}+\frac{i L_{x}}{2 v} \sum_{\sigma}\sum_{s}T_{\sigma_{a}\sigma}\frac{V_{\sigma s}V^{*}_{\sigma_{b}s}}{E-\epsilon_{s}}=\sum_{s}\frac{V_{\sigma_{a}s}V^{*}_{\sigma_{b}s}}{E-\epsilon_{s}}.
\end{equation}

To write down the solution of the above equation, we define the midgap level width due to coupling to the edge electrons, $\Gamma=\frac{L_{x}|V|^2}{2 v}$, as well as the energy difference between the incoming electron and the midgap level,  $\epsilon=E-\epsilon_{d}$. The T-matrix is then
\begin{equation}\label{eq:finalTmatrix}
    T\equiv\begin{pmatrix}
    T_{\uparrow\uparrow} & T_{\uparrow\downarrow} \\
    T_{\downarrow\uparrow} & T_{\downarrow\downarrow}
    \end{pmatrix}
    =
    \frac{|V|^{2}}{(\epsilon+i \Gamma)^{2}-b^{2}}
    \begin{pmatrix}
    \epsilon+i\Gamma+b \cos \theta & b \sin \theta e^{-i\phi} \\
    b \sin \theta e^{+i\phi} & \epsilon+i\Gamma-b \cos \theta
    \end{pmatrix}.
\end{equation}
\end{widetext}

Using the expression for the T-matrix, we can solve the scattering problem on the edge. To this end, we define two sets of scattering states with incoming waves incident from the left and right. These states $|\psi_\alpha\rangle$ are labeled with index $\alpha=L,R$, respectively, while the incident plane-wave states are denoted $|\psi_{in,\alpha}\rangle$. In position representation, the incoming electron wave functions  are given by 
\begin{equation}
    \braket{x|\psi_{in,L}}=
     \begin{pmatrix}
            e^{i k x} \\
            0 \\
        \end{pmatrix},
\end{equation}
\begin{equation}
    \braket{x|\psi_{in,R}}=
     \begin{pmatrix}
            0 \\
            e^{-ikx} \\
        \end{pmatrix},
\end{equation}
where $k=E/v$ is the momentum of the incoming electron. 

The scattering is characterized by transmission coefficients $t_{0}$ and $t_{0}^{'}$ and reflection coefficients $r_{0}$ and $r_{0}^{'}$. These are defined via the asymptotic behavior of $\braket{ x|\psi_\alpha}$ for $x>0$ and $x<0$ (since the scattering region is assumed to be confined to $x=0$). Specifically, we have 
form \par
\begin{equation}
    \braket{x|\psi_{L}}=
    \begin{cases}
        \begin{pmatrix}
            t_{0} e^{i k x} \\
            0 \\
        \end{pmatrix} & x>0 \\
         \begin{pmatrix}
             e^{i k x} \\
            r_{0} e^{- ikx} \\
        \end{pmatrix} & x<0\\
    \end{cases},
\end{equation}
and 
\begin{equation}
   \braket{x|\psi_{R}}=
    \begin{cases}
    \\
         \begin{pmatrix}
            r_{0}' e^{+ikx} \\
            e^{-ikx} \\
        \end{pmatrix} & x>0 \\
        \begin{pmatrix}
          0 \\
          t_{0}'e^{-ikx} \\
        \end{pmatrix} & x<0 
    \end{cases}.
\end{equation}

The Lippmann-Schwinger equation for $|\psi_\alpha\rangle$ that allows to relate the scattering amplitudes to the T-matrix is
\begin{align}
    \langle x\ket{\psi_{\alpha}}=\langle x\ket{\psi_{in,\alpha}}+\langle x|G_{edge}T\ket{\psi_{in,\alpha}}.
\end{align}
In position representation the edge electron Green's function~\eqref{eq:edgeGF} is given by
\begin{widetext}
\begin{equation}
    G_{edge}(x,y)=\begin{pmatrix}
        -\frac{iL_{x}}{v}e^{i \frac{E(x-y)}{v}} \theta(x-y) & 0 \\
        0 & -\frac{iL_{x}}{v}e^{ \frac{-iE(x-y)}{v}} \theta(y-x) \\
    \end{pmatrix},
\end{equation}
while the T-matrix is completely local for momentum-independent hybridization between the midgap and edge states: $T_{\sigma_a\sigma_b}(x,y)=T_{\sigma_a\sigma_b}\delta(x)\delta(y)$, where $T_{\sigma_a\sigma_b}$ is given by Eq.~\eqref{eq:finalTmatrix}.

The transmission and reflection amplitudes are found to be
\begin{align}
    t_{0}&=\frac{\epsilon^2+\Gamma^2-b^2-2ib\Gamma \cos\theta}{(\epsilon+i \Gamma)^2-b^2},\nonumber\\
    r_{0}&=\frac{-2 i \Gamma b \sin \theta e^{+i \phi}}{(\epsilon+i \Gamma)^2-b^2},\nonumber\\
    t_{0}'&=\frac{\epsilon^2+\Gamma^2-b^2+2ib\Gamma \cos\theta}{(\epsilon+i \Gamma)^2-b^2},\nonumber\\
    r_{0}'&=\frac{-2 i \Gamma b \sin\theta e^{-i \phi}}{(\epsilon+i \Gamma)^2-b^2}.
\end{align}
The transmission probability through a single midgap level is then given by 
\begin{align}\label{eq:transmissionmidgap}
    T_{mg}=|t_0|^2=|t'_0|^2=\frac{(\epsilon^2+\Gamma^2-b^2)^2+4b^2\Gamma^2\cos^2\theta}
    {(\epsilon^2+\Gamma^2-b^2)^2+4b^2\Gamma^2}.
\end{align}
\end{widetext}
It is apparent that $T_{mg}$ satisfies obvious conditions $T_{mg}(b=0)=T_{mg}(\theta=0)=1$.

To understand the significance of Eq.~\eqref{eq:transmissionmidgap}, let us consider the case of a magnetic field field perpendicular to the edge spin polarizations, $\theta=\pi/2$, and a resonant midgap level, $\epsilon\equiv E-\epsilon_d=0$. The dependence of the transmission coefficient on the magnitude of the Zeeman field, $b$, is shown in Fig.~\ref{fig:Timp}. It is apparent that the transmission is dramatically suppressed for $b\sim \Gamma$, and becomes close to unity for $b\ll\Gamma$ and $b\gg \Gamma$. This suppression for $b\sim \Gamma$ is crucial for the magnetoconductance of a sample with many impurities. 
\begin{figure}[htb]\label{fig:Timp}
\begin{minipage}{8.6cm}
\begin{tikzpicture}
  \node (img)  {\includegraphics[width=7.6cm]{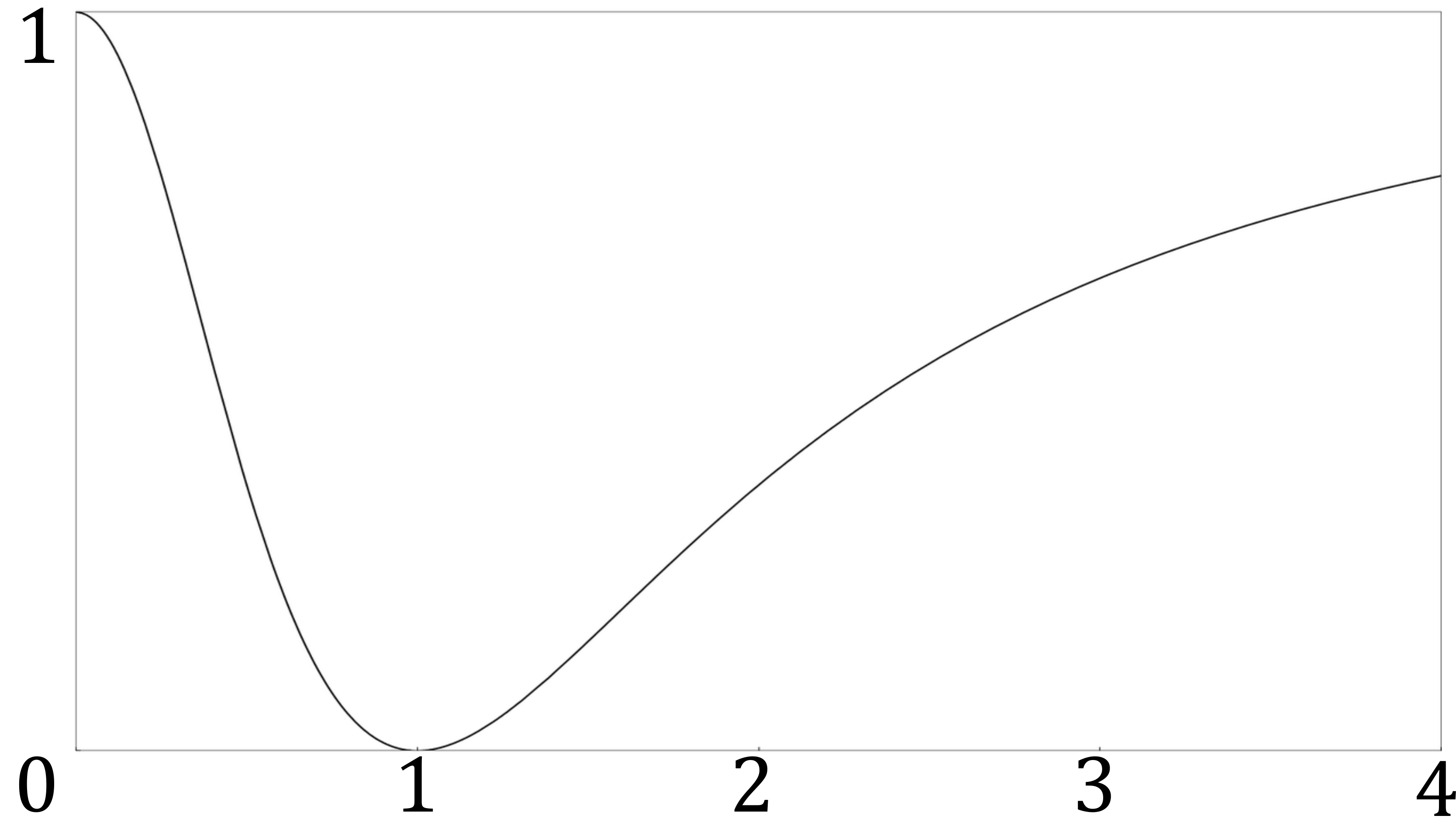}};
  \node[below=of img, node distance=0cm, yshift=1cm,font=\color{black}] {$b/\Gamma$ };
  \node[left=of img, node distance=0cm, rotate=90, anchor=center,yshift=-0.7cm,font=\color{black}] {$T_{mg}$};
 \end{tikzpicture}
\end{minipage}%
\caption{\label{fig:Timp}The transmission probability through a midgap level, $T_{mg}$, as a function of Zeeman field strength, $b$, measured in units of the level width, $\Gamma$, for $\theta=\pi/2$, $\epsilon=0$}
\end{figure}

The dependence of the transmission probability on the magnitude of the Zeeman field can be understood qualitatively as follows. As an electron gets scattered by a midgap level, the time it spends on the level is roughly $\Delta t\sim \min\left(\Gamma^{-1},\epsilon^{-1}\right)$. For small $b$, one can think about classical spin precession during the time spent on the level, which leads to spin rotation by $\Delta\phi\sim b\Delta t\sim b/\Gamma$ (for $\theta=\pi/2$). If $\Delta\phi$ becomes comparable to unity, the electron exiting the level can enter either of the helical states, since its wave function has considerable overlap with both spinors, and this results in considerable backscattering. For large enough fields, $b\gtrsim\Gamma$, the above semiclassical reasoning fails, and transmission is restored as the two Zeeman-split midgap levels decouple from the edge electrons. The overall conclusion of this argument is that midgap levels with $\epsilon\sim \Gamma\sim b$ can lead to strong backscattering of helical electrons even for small values of $b$.

These considerations have important implications for transport along an edge coupled to multiple midgap levels located in the bulk of the system. Under such circumstances, the level width varies with the real-space level distance from the edge. There is then a group of optimally located in real and energy spaces levels, which satisfy the $\epsilon\sim \Gamma\sim b$ condition, and provide strong backscattering even for small magnetic fields. We will describe this situation quantitatively in Section~\ref{sec:transport}.

We close this Section with a brief comparison of the transmission probability for a midgap level and the one for a potential impurity model, $T_{pot}$, which is considered in Appendix~\ref{sec:potentialscattering}. We quote here Eq.~\eqref{eq:potentialtransmission} for $T_{pot}$:
\begin{equation}\label{eq:Tpofromappendix}
    T_{pot}=\frac{E^2-b_{e}^2}{E^2-\left(\frac{4-\omega^2}{4+\omega^2}\right)^2b_{e}^2},
\end{equation}
where $b_e$ is the Zeeman energy for the edge electrons, defined just like $b$, but with g-factor of the midgap level replaced with the one for edge electrons, and dimensionless quantity $\omega$ is proportional to the strength of the impurity potential. See Appendix~\ref{sec:potentialscattering} for details. 

The edge electron states relevant for transport are those located near the Fermi level, such that the energy $E$ of the scattering electron is of the order of the spin splitting of the edge states at the Fermi level. It is then clear from Eq.~\eqref{eq:Tpofromappendix} that unless the Fermi level is within $\sim 0.1$meV of the Dirac point energy, the scattering on a potential impurity is very weak, and the corresponding transmission probability goes as $T_{pot}\propto b_e^2/E^2$.

\section{Transport on a helical edge with many impurities}\label{sec:transport}

\begin{figure}
    \centering
    \includegraphics[width=3.5in]{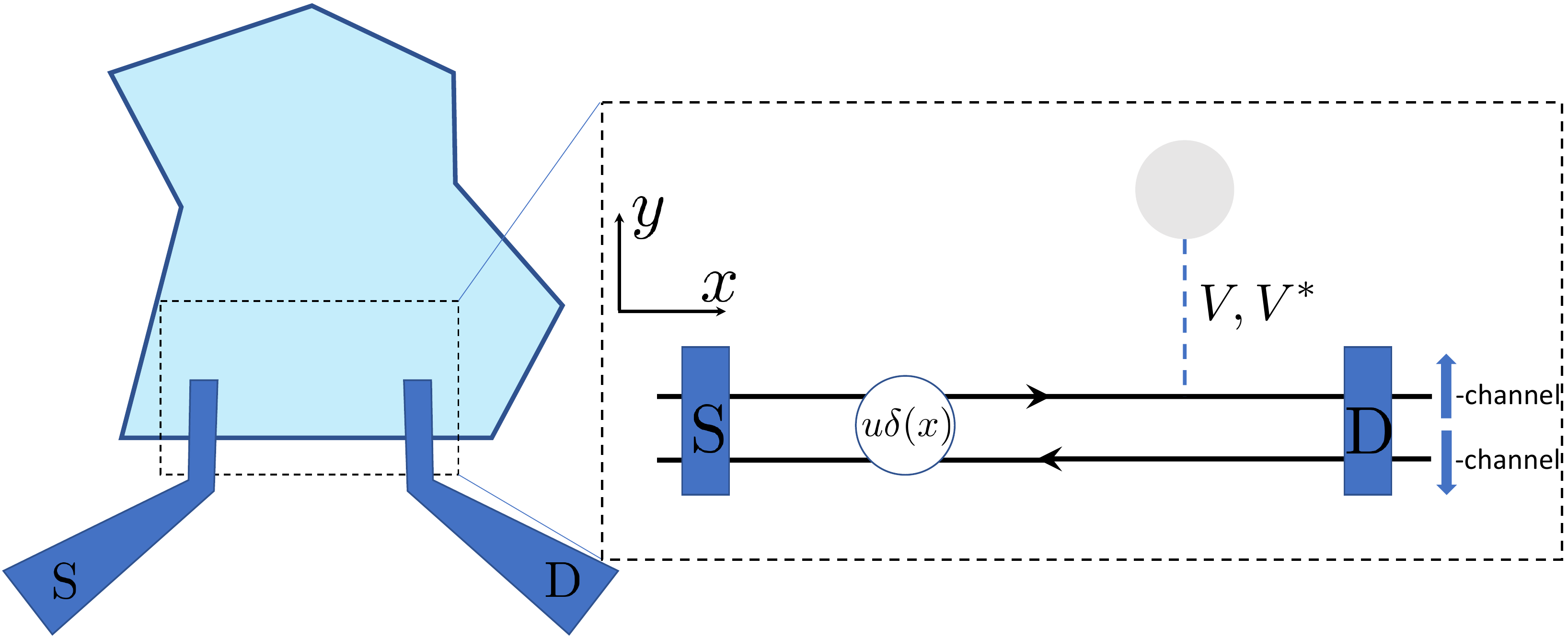}
    \caption{(Color online) Left: Typical experimental setup for a flake of WTe$_2$, with source (S) and drain (D) electrodes attached to a particular edge. Right: schematic view of the shortest edge confined between the source and drain electrodes, with two types of disorder for helical electrons traveling along the edge. The empty circle depicted right on top of the helical edge states represents a potential impurity with potential $u\delta(x)$, centered at the location of the impurity. The shaded circle is a bulk midgap state, hybridized with the helical edge states by a matrix element $V_p$.}
    \label{fig:flake}
\end{figure}

In this Section we consider transport on a topological edge in the presence of multiple impurities. Realistic experiments~\ on WTe$_2$ are often performed on irregular flakes, with contacts placed like shown schematically in Fig.~\ref{fig:flake}. It typically observed~\cite{Pesin_2020} that even the zero-B-field two-probe conductance in long channels is far from its quantized value. This makes it natural to assume that the two-probe conductance between the source and drain electrodes is determined by backscattering in the shortest part of the edge confined between them. We restrict our attention to this single portion of the entire edge, assuming that the complementary part is long enough to neglect electron transfer between the source and drain along it. 
To describe the short-channel set-up, like the one in Ref.~\cite{wu2018wte2}, one just needs to include the transport along the complementary part of the sample's edge, which presents no difficulty, if the inter-edge scattering is absent. 

Under the realistic circumstances, the transport along a helical edge is partially coherent due to finite temperature, electron-electron as well as electron-phonon interaction, and possibly other factors. Our goal is to show that weakly side-coupled midgap bulk states lead to a cusp-like behavior of the magnetoconductance at small B-fields. Hence we mostly disregard the electron-electron and electron-phonon interaction on the edge. (This question was considered, for instance, in Refs.~\cite{schmidt2012inelastic,mirlin2014conductivity}.) We consider two limits of edge transport: fully coherent, in which backscatttering on the edge leads to localization, and fully incoherent transport, assuming there are local chemical potentials for each species of the helical electrons. 

\subsection{Coherent transport: Localization \label{sec:coherenttransport}}
We assume that the potential impurities on the edge and the midgap states are dilute enough, such that each of them can be considered as an independent scatterer. The potential impurities are assumed to be all identical to each other, and  are described with a linear density $n_p$. The midgap states have a two-dimensional density in the real space, $n_{mg}$, and their energies, $\epsilon_{mg}$, are distributed with certain probability density $\rho(\epsilon_{mg})$ in the energy space, normalized to unity. We will make a realistic assumption that $\rho(\epsilon_{mg})$ does not have much structure as a function of $\epsilon_{mg}$ on the scale of the Zeeman energy. That is, we will take $\rho(\epsilon)\equiv\rho$, but the value of the constant $\rho$ probably depends on the gate voltage.

For the fully coherent case, the logarithm of the transmission coefficient is an additive self-averaging quantity~\cite{datta2005quantum}. Assuming that the edge has length $L_x$, and interacts with midgap states from a rectangular region of width $L_y$ ($L_y$ will cancel out from all final results), and that the distribution of the number of the impurities and midgap states is Poissonian, we can write the following expression for the disorder-averaged transmission probability on the edge:
\begin{equation}
    \ln T_{tot}= n_pL_x \ln T_{pot} + n_{mg}L_xL_y \langle\ln T_{mg}\rangle. 
\end{equation}
In the equation above $T_{pot}$ is the transmission probability of a potential impurity, Eq.~\eqref{eq:potentialtransmission}.  $\langle\log T_{mg}\rangle$ is the disorder-averaged logarithm of the transmission coefficient due to a midgap state, calculated below.  

Our main goal is to determine the behavior of the magnetoconductance at small fields, and to extract its singular part. It is clear from Eq.~\eqref{eq:potentialtransmission} that the transmission coefficient through a potential impurity admits a regular expansion in powers of $b_e^2/E^2\propto B^2\sin^2\theta$ at a finite doping level on the edge, so the singular part can only come from the midgap states. Hence we focus on $\langle\log T_{mg}\rangle$ below, neglecting the effects associated with $T_{pot}$.

It follows from Eq.~\eqref{eq:transmissionmidgap} that the transmission coefficient due to a midgap state can be written as 
\begin{align}
    T_{mg}=T_{mg}(\epsilon,\Gamma,\theta), 
\end{align}
where, again, $\epsilon$ is the difference between the energy of the electron and the energy of the midgap state, $\Gamma$ is the midgap state width due to coupling to the helical edge, and $\theta$ is the angle between the direction of the magnetic field, and the edge spin polarization in the absence of the field. (It does not matter what helicity is taken to define the positive direction, since the angular dependence is invariant with respect to $\theta\to \pi-\theta$.) As before, we disregard the effect of the Zeeman field on the edge spin polarization at small magnetic fields while calculating transmission through a midgap state, since it does not affect the small-field behavior of the conductance. 

To perform averaging over disorder, we assume a simple model in which the midgap level width depends on its distance to the edge, $\Gamma=\Gamma(y)$.  Then impurity averaging over the midgap level spatial position and energy is defined as 
\begin{align}\label{eq:impurityaveraging}
    \langle\ldots\rangle=\frac{\rho}{ L_y}\int^{L_y}_0 dy \int^{\infty}_{-\infty}d\epsilon_{mg} (\ldots)
\end{align}
The limits for the energy integral are set to be infinite under the assumption that the quantity being averaged will provide convergence of the integral. 

We will perform the integration over $y$ with
\begin{align}
    \Gamma(y)=\Gamma_0 e^{-y/a},
\end{align}
which is borrowed from the form of overlap integral between localized impurity states in the theory of doped semiconductors~\cite{efros}. The length scale $a$ roughly coincides with the spatial extent of the midgap level wave function, and $\Gamma_0$ is the width of the level located very close to the edge. We expect that $\Gamma_0\gg b$, since it originates from unit-cell scale physics. 

Performing the integrals in Eq.~\eqref{eq:impurityaveraging} with the transmission coefficient from Eq.~\eqref{eq:transmissionmidgap}, we obtain in the $\Gamma_0\to\infty$ limit
\begin{equation}\label{eq:averagelogT}
    \langle\ln T_{mg}\rangle\approx  -2\pi^2\rho \frac{a}{L_y}b(1-|\cos\theta|). 
\end{equation}
For a finite $\Gamma_0$, corrections to this result go as $b^2/\Gamma_0$, and are expected to be very small. 

Eq.~\eqref{eq:averagelogT} is one of the main results of this work. The notable features are its linear scaling with the magnitude of the Zeeman field, $b$, and its dependence on the angle $\theta$ between the direction of the magnetic field, and the edge spin polarization in the absence of a magnetic field. Since we have no reliable estimates for the prefactor in the right hand side of Eq.~\eqref{eq:averagelogT}, the dependence on $b$ and $\theta$ should be considered as the main signatures of this mechanism of magnetoconductance. Note that while linear dependence on $b$ can be expected for chemical potentials near an edge Dirac point due to opening of an edge gap, the angular dependence in that case would be given by $|\sin\theta|$, which can be distinguished from $1-|\cos\theta|$ given the precision of existing experiments. Of course, an activated mechanism of conductance can be distinguished from the present one by its temperature dependence, not just angular one. 

Using Eq.~\eqref{eq:averagelogT}, we can write down the expression for the magnetoconductance in the coherent regime if we assume that other mechanisms of backscattering are not very sensitive to weak magnetic fields. For instance, backscattering due to magnetic impurities requires magnetic anisotropy for the latter\cite{matveev2011kondo,aleiner2013localization}, and one expects that weak magnetic fields will not overpower that anisotropy to change the conductance appreciably. Restoring the magnetic field magnitude, $B$, in place of the Zeeman energy scale $b$, we thus obtain for the edge conductance $G(B)$ at weak fields in the coherent (localization) regime:
\begin{align}\label{eq:conductancecoherent}
    \frac{G(B)}{G(0)}=\exp\left[\ln T_{tot}\right]\approx \exp\left[-\frac{L_x}{\ell}\frac{|B|}{B_0}(1-|\cos\theta|)\right],
\end{align}
where $\ell=1/n_{mg}a$, $B^{-1}_0=\pi^2\rho g_{mg}\mu_B$. We expect $B_0$ to be a rather large field scale, since it roughly corresponds to the Zeeman energy of $\rho^{-1}$, which is probably at least similar to the band gap. The overall factor of $(L_x/\ell)(B/B_0)$ is the number of midgap states located within a real-space strip of width $a$ near the edge of the system, and energy-space strip of width $\sim g_{mg}\mu_B B$ near the Fermi level. Eq.~\eqref{eq:conductancecoherent} is applicable for magnetic fields that are not too small, for which $(L_x/\ell)(B/B_0)\gtrsim 1$. 

We will further discuss Eq.~\eqref{eq:conductancecoherent} in Section~\ref{sec:discussion}, turning now to the incoherent edge transport.

\subsection{Incoherent transport\label{sec:incoherenttransport}}

In this Section, we consider the regime in which there is inelastic relaxation on the edge, strong enough to establish (separate) local equilibria on each of the two helical channels. This regime should be relevant at high enough temperatures, such that inelastic processes due to electron-electron and electron-phonon interaction are frequent enough.

\begin{figure}
    \centering
    \includegraphics[width=3.5in]{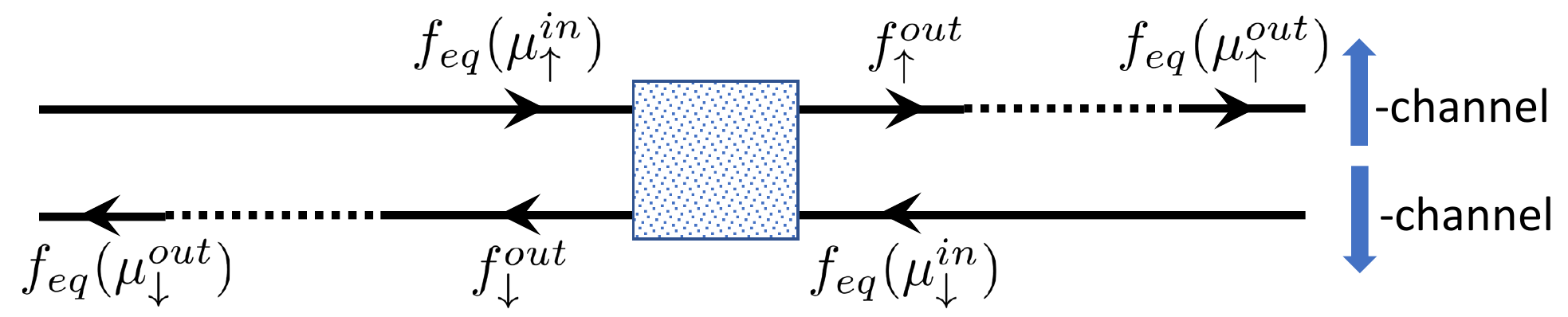}
    \caption{Schematic representation of edge transport in the incoherent regime. The shaded rectangle represents the scattering region. Also shown the distribution functions of the incoming (equilibrium) and outgoing states right after scattering (out of equilibrium). The distribution functions of the outgoing states are relaxed to equilibrium ones with channel-specific chemical potentials away from the scattering region, before the next scattering event.}
    \label{fig:incoherent}
\end{figure}

An elastic scattering event, regardless of the type of impurity causing it, followed by inelastic relaxation to a local-distribution with nonequilibrium chemical potential temperature is shown schematically in Fig.~\ref{fig:incoherent}. 
Each elastic scattering event, occurring at a given energy, connects the electron distribution functions of the outgoing electrons to those of incoming ones, assumed to have the equilibrium form. We will denote quantities pertaining to the incoming and outgoing electrons with superscripts `\textit{in}', and `\textit{out}', respectively. Furthermore, we will use subscripts $\uparrow,\downarrow$ to indicate to which helical channel - up-spin or down-spin - various quantities pertain. We will use $T(E)$ and $R(E)=1-T(E)$ for the elastic transmission and reflection coefficients, without specifying the type of the impurity, since the considerations are general. With this notation, we obtain the following equations for the distribution functions of electrons involved in an act of elastic scattering: 
\begin{align}
    &f^{out}_{\uparrow}=T(E) f_{eq}(\mu^{in}_\uparrow)+R(E)f_{eq}(\mu^{in}_\downarrow),\nonumber\\
    &f^{out}_{\downarrow}=T(E) f_{eq}(\mu^{in}_\downarrow)+R(E)f_{eq}(\mu^{in}_\uparrow),
\end{align}
where $f_{eq}(\mu)=[1+e^{\beta(E-\mu)}]^{-1}$ is the Fermi-Dirac distribution function with inverse temperature $\beta$.
The values of the chemical potentials for the outgoing electrons are obtained from the condition that inelastic scattering within each helical channel conserves the number of electrons in that channel, leading to the condition 
\begin{align}
    \int dE \nu_\sigma(E) f^{out}_{\sigma}=\int dE \nu_\sigma(E) f_{eq}(\mu^{out}_\sigma),
\end{align}
with $\sigma=\uparrow,\downarrow$, and $\nu_\sigma(E)$ being the density of states for the two helical channels. If these densities of states are approximately constant near the Fermi level, and deviations from equilibrium are small, we immediately obtain 
\begin{align}\label{eq:chempots}
    \mu^{out}_\uparrow=\bar{T}\mu^{in}_{\uparrow}+\bar{R}\mu^{in}_{\downarrow},\nonumber\\
    \mu^{out}_\downarrow=\bar{T}\mu^{in}_{\downarrow}+\bar{R}\mu^{in}_{\uparrow}.
\end{align}
In Eq.~\eqref{eq:chempots} we introduced energy-averaged transmission and reflections coefficients: 
\begin{align}\label{eq:energyaverage}
\bar{T}=\int dE\, T(E)\left(-\frac{\partial f_{eq}}{\partial E}\right), \quad \bar{R}=1-\bar{T}.
\end{align}

Eqs.~\eqref{eq:chempots} allow us to calculate the intrinsic, or ``four-probe", resistance of the impurity in question. Denoting this resistance as the inverse of the corresponding conductance, $G^{-1}_{imp}$, since we reserved $R$ for the reflection probability in an impurity problem, we obtain~\cite{datta2005quantum} 
\begin{align}\label{eq:impurityresistance}
    G^{-1}_{imp}=\frac{h}{e^2}\frac{\bar{R}}{\bar{T}}.
\end{align}

We can now use Eq.~\eqref{eq:impurityresistance} to determine magnetoconductance on an edge with many impurities. To account for conductance suppression in zero magnetic field, the theory of which we are not trying to build at all, we assume that various scattering mechanisms obey the Mathiessen rule in the incoherent regime, such that the resistances are additive, and the inverse conductance on the edge can be  written as 
\begin{align}\label{eq:resistanceexact}
    G^{-1}(B)=G^{-1}(0)+\frac{h}{e^2}\sum_{imp}\frac{\bar{R}_{imp}}{\bar{T}_{imp}}.
\end{align}
The sum over impurities, $\sum_{imp}$, just adds together intrinsic resistances of all types of disorder present, while the quantum resistance of the helical channel itself, $h/e^2$, is included in the zero-field resistance $G^{-1}(0)$. 

We perform disorder averaging for the additive resistance~\eqref{eq:resistanceexact}, which amounts to averaging each impurity's intrinsic resistance. As before, we are interested in the small-field part of the magnetoconductance, and we expect the midgap states to provide a singular contribution. Therefore, we replace the sum over impurities by just the sum over midgap levels hybridized with the helical edge, $\sum_{imp}\to\sum_{mg}$. Then we obtain, much as in Section~\ref{sec:coherenttransport}:

\begin{equation}\label{eq:averageresistance}
    \sum_{mg}\left\langle\frac{\bar{R}_{mg}}{\bar{T}_{mg}}\right\rangle\approx n_{mg} L_{x} \rho \int_{-\infty}^{\infty}d \epsilon_{mg} \int_{0}^{\infty} dy \frac{\bar{R}_{mg}}{\bar{T}_{mg}}.
\end{equation}
We then notice that since $R_{mg}$ vanishes for $B=0$, at small B-fields we can set $T_{mg}\to1$ to obtain the leading in $B$ contribution to the resistance. After this step, the averaging is done for the reflection probability only. Since all integrals are converging, we can interchange the integrations over $\epsilon_{mg}$ and $E$ in Eq.~\eqref{eq:averageresistance}. In other words, energy averaging, Eq.~\eqref{eq:energyaverage}, and impurity averaging commute for $R_{mg}$. Because of our assumption that the midgap levels are distributed uniformly on energy scales appearing in the electron scattering problem, integration over $\epsilon_{mg}$ in Eq.~\eqref{eq:averageresistance} removes the dependence on the electron energy $E$ in the disorder-averaged reflection probability, hence energy averaging becomes trivial. Finally, performing all the integration, and again taking the $\Gamma_0/b\to\infty$ limit, we obtain
\begin{equation}
     \sum_{mg}\left\langle\frac{\bar{R}_{mg}}{\bar{T}_{mg}}\right\rangle\approx \pi^2 n_{mg} L_{x} \rho  a\,b\sin^2 \theta \equiv \frac{L_x}{\ell}\frac{|B|}{B_0}\sin^2\theta.
\end{equation}
The expression for the magnetoconductance in the incoherent regime then becomes 
\begin{align}\label{eq:conductanceincoherent}
    G(B)=\frac{1}{G^{-1}(0)+\frac{h}{e^2}\frac{L_x}{\ell}\frac{|B|}{B_0}\sin^2\theta}.
\end{align}
This is another important result of this work. It is clear that from the Taylor series in $B$ point of view this expression~\eqref{eq:conductanceincoherent} is very similar to Eq.~\eqref{eq:conductancecoherent}. Yet the two expressions can be distinguished by respective angular dependence. Again, the $\sin^2\theta$ dependence in the incoherent case is quite distinct from the $|\sin\theta|$ in the special case of the Fermi level being very close to an edge Dirac point, and the ensuing thermally-activated edge transport.

\section{Relation to experiment and discussion}\label{sec:discussion}
In this work we have studied how the presence of midgap states in the bulk of a quantum spin Hall insulator affects magnetotransport along its edge. We showed that hybridization of the midgap levels with the helical edge states leads to backscattering of the latter in the presence of a magnetic field. The backscattering probability can be of order unity for resonant midgap levels located not too close to the edge in real space, such that their width due to coupling to the helical edge states is comparable to the Zeeman coupling energy scale \emph{for the midgap level}. 

We showed that the presence of bulk midgap states leads to nonanalytic - cusp-like - dependence of the edge conductance on the magnetic field in both coherent and incoherent transport regimes. The two regimes can be distinguished by the angular dependence of the magnetoconductance, see Eqs.~\eqref{eq:conductancecoherent} and~\eqref{eq:conductanceincoherent}.

Given specific angular dependencies obtained in this work, it is tempting to compare them to those observed experimentally in monolayer WTe$_2$. This material belongs to the $C^2_{2h}$ crystallographic class, and has a mirror plane perpendicular to the monolayer. It was previously argued~\cite{thomale2019prb,Pesin_2020,garcia2020canted} that the antiparallel spin polarizations of the counterpropagating helical edge states are determined by the bulk band structure (rather than the details of the edge in a particular sample, or gate voltage for the same sample), and lie in the aforementioned mirror plane. We will denote the angle that the spin polarization of one of the helical channel makes with the normal to the sample as $\theta_0$. It does not matter which channel's spin polarization is chosen to define $\theta_0$, since all the results are invariant with respect to $\theta_0\to \pi-\theta_0$. 

In Ref.~\cite{Pesin_2020} the magnetoconductance of monolayer WTe$_2$ was studied as a function of the magnetic field magnitude, and its direction in the mirror plane of the sample. Below we will denote $\theta$ to be angle that the magnetic field makes with the normal to the monolayer, rather than with the edge spin polarization direction, as in the preceding part of the paper. 
In Ref.~\cite{Pesin_2020}, the direction of the spin polarization of the edge electrons was determined as the direction of the magnetic field for which the suppression of the edge conductance was the smallest, but still non-zero. 

To compare the present theory to the experimental data, we note that in both coherent and incoherent transport cases the sample resistance, $R(B,\theta)$, is predicted to rise linearly with the magnetic field, with some rounding at the smallest fields (see the discussion at the end of the Introduction). Neglecting the rounding, we expect the resistance to go as 
\begin{align}
    R(B,\theta)=R(0,\theta)[1+\beta_1(\theta) |B|],
\end{align}
in both regimes of transport, see Eqs.~\eqref{eq:conductancecoherent} and~\eqref{eq:conductanceincoherent}.

The specific prediction of the theory is the angular dependence of the coefficient $\beta_1(\theta)$, which is given by 
\begin{align}
    &\beta_1^{coh}\propto (1+|\cos(\theta-\theta_0)|),\nonumber\\
    &\beta_1^{incoh}\propto \sin^2(\theta-\theta_0),
\end{align}
for the coherent and incoherent regimes, respectively. These angular dependencies are quite distinct from each other, as well as from the case of the chemical potential being in the edge gap, for which the angular dependence of $\beta_1$ should follow the $|\sin(\theta-\theta_0)|$ form.

\begin{figure}
    \centering
    \includegraphics[width=3.5in]{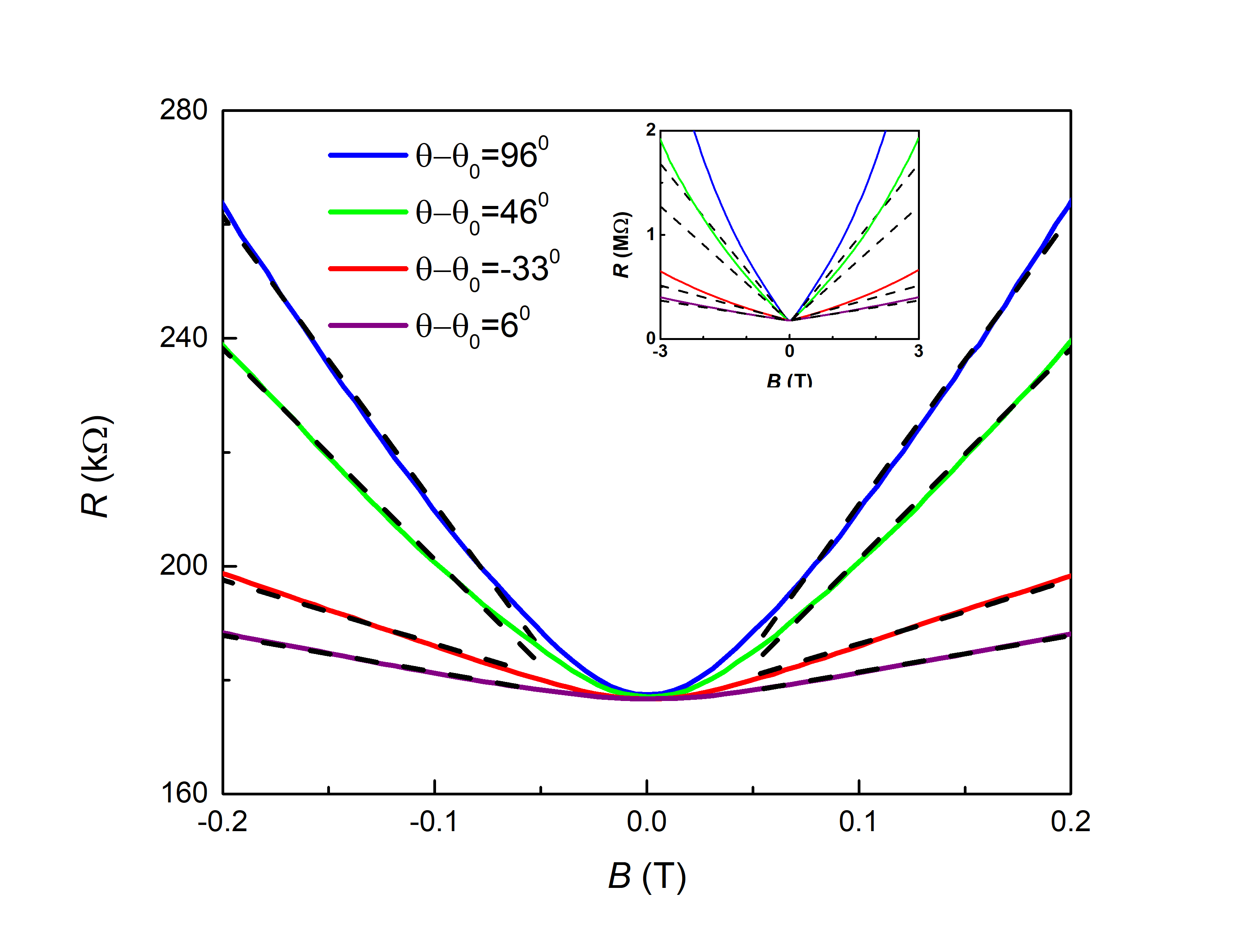}
    \caption{(Color online) Fits of experimental data at low magnetic fields ($0.05T<B<0.2 T$) to linear dependence (main panel) at various field angles. The inset demonstrates deviations from the small-field linear magnetoresistance at larger values of $B$.}
    \label{fig:data}
\end{figure}

\begin{figure}
    \centering
    \includegraphics[width=3.5in]{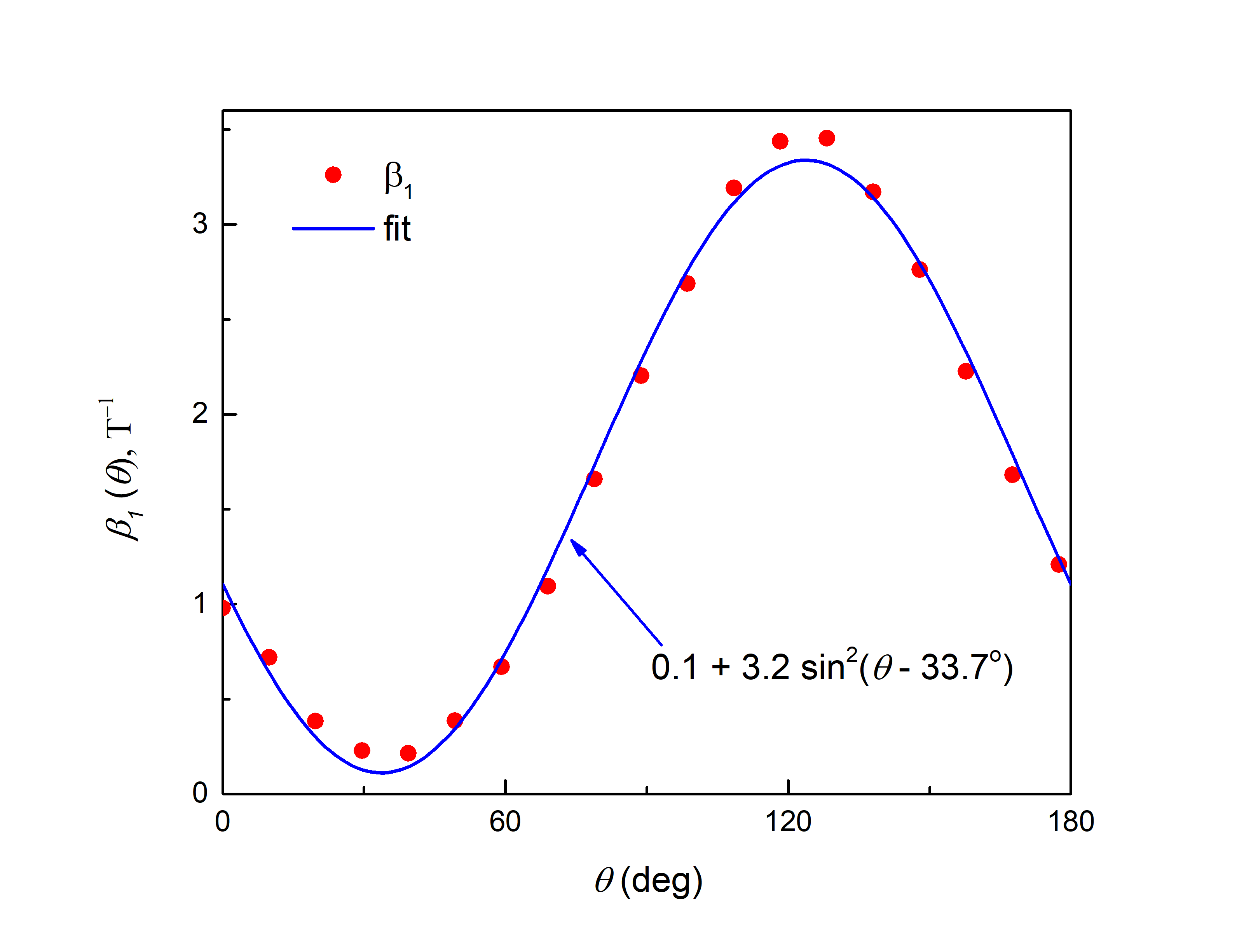}
    \caption{(Color online) Experimentally extracted slope of resistance at small magnetic fields ($0.05T<B<0.2 T$) as a function of the field angle (red dots). The continuous blue line is the least-squares fit to $\beta_1(\theta)=c_0+c_1\sin^2(\theta-\theta_0)$, with $c_0=0.1 \rm{T}^{-1},c_1=3.2 \rm{T}^{-1}$ and $\theta_0=33.7^\circ$.}
    \label{fig:beta}
\end{figure}

Fitting the experimental data for the sample resistance to a linear function at small magnetic fields, we first of all observe that linear-in-B dependence describes the small-field magnetoresistance well, see Fig.~\ref{fig:data}. Furthermore, we can extract the angular dependence of the observed $\beta_1(\theta)$ from the fits, which is presented in Fig.~\ref{fig:beta}. It turns out that 
\begin{align}
    \beta_1(\theta)=c_0+c_1\sin^2(\theta-\theta_0)
\end{align}
with $c_0= 0.1\,\textrm{T}^{-1}$, $c_1= 3.2\,\textrm{T}^{-1}$ fits the values extracted from the experimental data very well, see Fig.~\ref{fig:beta}. This suggests that the edge transport is described with the incoherent regime of the present theory.

The phenomenological constant term $c_0$ obtained from the data can possibly stem from $\theta_0$ defining only the average direction of the edge spin polarization, the variation of which can come from random spin-orbit coupling due to the local electric fields near the edge, or anisotropies of the midgap state $g$-factor. Under such circumstances, we expect that there will be finite backscattering even for a magnetic field aligned with the average edge spin polarization, which is captured by $c_0\neq0$.

The presented analysis suggests that the transport in a few-micron-long edge channels of monolayer WTe$_2$ is largely incoherent at T=4K. It would be interesting to study the dependence of the shape of $\beta_1(\theta)$ on temperature. Such $\beta_1(\theta,T)$ could shed light on the coherent-incoherent transition in the quantum spin Hall insulator edge transport. 

\begin{acknowledgments}
We are grateful to Anton Andreev, Leonid Glazman, Igor Gornyi, and Mikhail Raikh for useful discussions. The theoretical work was supported by the National Science Foundation Grant No. DMR-2138008 and experimental work by NSF DMR grant MRSEC 1719797.
\end{acknowledgments}
\appendix

\section{Helical edge with a potential impurity\label{sec:potentialscattering}}
We consider an infinitely long one-dimensional system of helical fermions with a single short-range impurity. We model this problem with a Hamiltonian that consists of three parts: 
\begin{equation}
    H_{edge}=H_{0}+H_{Z}+H_{imp},
\end{equation}
where $H_{0}$ is the Hamiltonian of edge electrons with their spin polarization defining the $z$-direction. $H_{Z}$ is the Zeeman coupling term between external magnetic field in the $x$-direction. $H_{imp}$ is the potential impurity on edge. The three terms in the Hamiltonian are as follows. First, 
\begin{equation}\label{eq:H0}
    H_{0}=\sum_{\sigma\sigma'}\int_{-\infty}^{+\infty} dx
    \psi_{\sigma}^{\dagger}(x)
    \left[-i v\sigma_z \partial_{x}\right]_{\sigma\sigma'}
    \psi_{\sigma'}(x),
\end{equation}
where $\psi_\sigma(x)$ is the field operator annihilating an electron with spin projection $\sigma$ at spatial position $x$, and $v$ is the Dirac speed, and $\sigma_z$ is a Pauli matrix. 
\begin{equation}\label{eq:HZeeman}
    H_{Z}=\sum_{\sigma\sigma'}\int_{-\infty}^{+\infty} dx
    \psi_{\sigma}^{\dagger}(x)
    \left[b_e \sigma_{x}\right]_{\sigma\sigma'}
    \psi_{\sigma'}(x),
\end{equation}
To simplify the notation, we describe the magnetic field with the corresponding Zeeman energy, $b_e=\frac12g_{e}\mu_B B$, 
where $g_e$ is the $g$-factor for the edge electrons, $\mu_B>0$ is the Bohr magneton, and $B$ is the magnetic induction. 

Finally, 
\begin{equation}
    H_{imp}=\sum_{\sigma}\int_{-\infty}^{+\infty}dx  U(x)  \psi^{\dagger}_{\sigma}(x)\psi_{\sigma}(x),
\end{equation}
in which we will take the impurity potential $U(x)$ to be represented by a $\delta$-function for simplicity: $U(x)=u \delta(x)$, where $u$ describes the impurity strength.

In this section, we will use the Lippmann-Schwinger equation to solve the transmission and reflection probability of potential scattering. We use the helical edge Hamiltonian ($H_{helical}$) plus the Zeeman coupling term ($H_{mag}$) as the unperturbed Hamiltonian. In second quantization form, the unperturbed Hamiltonian is 
\begin{equation}
    H_{free}=\sum_{\sigma\sigma'}\int_{-\infty}^{+\infty} dx
    \psi_{\sigma}^{\dagger}(x)
    \left[-i v\sigma_z \partial_{x}+b_{e} \sigma_{x}\right]_{\sigma\sigma'}
    \psi_{\sigma'}(x).
\end{equation}
 In momentum space, the unperturbed Hamiltonian can be written as \par
 \begin{equation}
     H_{free}=\sum_{p}
     \begin{pmatrix}       a^{\dagger}_{k,\uparrow} & a^{\dagger}_{k,\downarrow}\\
     \end{pmatrix}
       \begin{pmatrix}       vp & b_{e} \\
       b_{e} & -vp \\
     \end{pmatrix}
      \begin{pmatrix}       a_{k,\uparrow}\\
a_{k,\downarrow} \\
     \end{pmatrix}.
 \end{equation}
\begin{widetext}
The energy spectrum for this Hamiltonian is $E_{spectrum}=\pm \sqrt{v^2k^2+b_{e}^{2}}$. We assume that the energy of the incoming electron is positive, i.e $E>0$. The energy for conduction band is $E_{c}(k)=\sqrt{v^2k^2+b_{e}^2}$ and the energy band for valence band is $E_{v}(k)=-\sqrt{v^2k^2+b_{e}^{2}}$. In this case the electron can be left-propagating or right-propagating. The right-propagating electron eigenstate $\ket{\chi_{c}(k_{+})}$ has momentum $k_{+}=\frac{\sqrt{v^2k^2+b_{e}^{2}}}{v}$ ,while the left-propagating electron eigenstate $\ket{\chi_{c}(k_{-})}$ has momentum $k_{-}=-\frac{\sqrt{v^2k^2+b_{e}^{2}}}{v}$. The expressions for eigenstates are \par
\begin{equation}
    \left|\chi_{c}(k_{+})\right\rangle =\frac{1}{\sqrt{(E-\sqrt{E^2-b_{e}^2})^{2}+b_{e}^{2}}}
    \begin{pmatrix}
    b_{e}\\
    E-\sqrt{E^2-b_{e}^2} \\
    \end{pmatrix},
\end{equation}
and
\begin{equation}
    \left|\chi_{c}(k_{-})\right\rangle=\frac{1}{\sqrt{(E-\sqrt{E^2-b_{e}^2})^{2}+b_{e}^{2}}}
    \begin{pmatrix}
    E-\sqrt{E^2-b_{e}^2} \\
    b_{e} \\
    \end{pmatrix}.
    \end{equation} \par 
The  Green's function for the helical edge with Zeeman coupling term between external magnetic in $x$ direction field and the edge  is\par 
\begin{equation}
    G(E+i \eta)=\sum_{k}\sum_{ss' }\frac{\ket{ks}\bra{ks'}}{E+i \eta- H_{free}}=\sum_{k}\sum_{ss'}\frac{\ket{ks}(E+i\eta + v k \sigma_{z}+b_{e}\sigma_{x})_{ss'}\bra{ks'}}{(E+i \eta)^{2}-v^2 p^2-b_{e}^2},
\end{equation}
Where k is the momentum of the electron and $ss'=\uparrow$ or $\downarrow$ are the spin indices of the electron wavefunction.\par 
The potential impurity can also be written in the basis of spin and momentum of electrons. In momentum space the interaction between state with momentum k and spin s and state with momentum $k'$ and spin $s'$ is uniform $ V^{k k'}_{s s'}=\bra{k s} V \ket{k's'} =\frac{u}{L_{x}} \delta_{s s'}$, where $L_{x}$ is the length of the edge. so the V matrix is \par 
\begin{equation}
    V= \sum_{k,s}\sum_{k',s'}\ket{k' s'}V^{k k'}_{s s'} \bra{ks}=\sum_{k k'} \sum_{s }\ket{ks}\frac{u}{L_{x}} \bra{k's} .
\end{equation}
The T-matrix equals the sum of products of Green's Functions and interactions: $T=(\sum_{n=0}^{\infty}(VG)^{n})V=(1-VG)^{-1}V$ , so we can derive T-Matrix using the Identity $T-VGT=V$. Defining dimensionless constant $\omega=\frac{u}{v}$, we then use the definition $T^{ k_{a} k_{b}}_{ \sigma_{a}\sigma_{b}}=\bra{k_{a}\sigma_{a}} T \ket{k_{b}\sigma_{b}}$ to describe the scattering matrix between edge state $\ket{k_{a}\sigma_{a}}$ and $\ket{k_{b}\sigma_{b}}$. We get the algebraic equation for the T-matrix\par
\begin{equation} T^{k_{a}k_{b}}_{\sigma_{a}\sigma_{b}}-\frac{u}{2\pi}\sum_{s}\int dk \frac{(E+i\eta+vk\sigma_{z}+b_{e}\sigma_{x})_{\sigma_{a}s}T^{kk_{b}}_{s\sigma_{b}}}{(E+i\eta)^2-v^2 k^2-b_{e}^{2}}=\frac{u}{L_{x}}\delta_{\sigma_{a}\sigma_{b}}.
\end{equation}
Since the V matrix is independent of momentum, we can assume that the T-matrix is also independent of momentum. So the equation for the T-matrix can be simplified into \par
\begin{equation} T_{\sigma_{a}\sigma_{b}}-\frac{u}{2\pi}\sum_{s}\int dk \frac{(E+i\eta+vk\sigma_{z}+b_{e}\sigma_{x})_{\sigma_{a}s}T_{s\sigma_{b}}}{(E+i\eta)^2-v^2 k^2-b_{e}^{2}}=\frac{u}{L_{x}}\delta_{\sigma_{a}\sigma_{b}}.
\end{equation}
Integrating out the internal momentum k, we will get an algebraic equation for the  T-matrix.\par
\begin{equation}
    T_{\sigma_{a}\sigma_{b}}+\sum_{s}\frac{ i \omega}{2} \frac{(E+b_{e}\sigma_{x})_{\sigma_{a}s}T_{s\sigma_{b}}}{\sqrt{E^2-b_{e}^{2}}}=\frac{u}{L_{x}}\delta_{\sigma_{a}\sigma_{b}}.
\end{equation}
Solving the algebraic equation for T-matrix, we get the expression for T-matrix.\par
\begin{equation}
    T=\frac{u/L_{x}}{(1+\frac{i \omega E}{2 \sqrt{E^{2}-b_{e}^{2}}})^{2}+(\frac{\omega b_{e}}{2 \sqrt{E^{2}-b_{e}^{2}}})^{2}}
    \begin{pmatrix}
    1+\frac{i \omega E}{2 \sqrt{E^{2}-b_{e}^{2}}} & -\frac{ i\omega b_{e}}{2 \sqrt{E^{2}-b_{e}^{2}}} \\
    -\frac{i\omega b_{e}}{2 \sqrt{E^{2}-b_{e}^{2}}} & 1+\frac{i \omega E}{2 \sqrt{E^{2}-b_{e}^{2}}} \\
    \end{pmatrix}.
\end{equation}
The real space Green's function for helical edge with Zeeman coupling term between external magnetic field in x direction is
\begin{equation}
    G(x,y)=\frac{L_{x}}{2\pi} \int dk \frac{e^{ik(x-y)}}{E+i \eta -H_{free}}e^{ik(x-y)}
    =
    \begin{cases}
        -\frac{i L_{x}}{2 v} \frac{E+\sqrt{E^2-b_{e}^{2}}\sigma_{z}+b_{e}\sigma_{x}}{\sqrt{E^2-b_{e}^{2}}}e^{\frac{i \sqrt{E^2-b_{e}^2}(x-y)}{v}} & x>y \\
        -\frac{i L_{x}}{2 v} \frac{E-\sqrt{E^2-b_{e}^{2}}\sigma_{z}+b_{e}\sigma_{x}}{\sqrt{E^2-b_{e}^{2}}}e^{-\frac{i \sqrt{E^2-b_{e}^2}(x-y)}{v}} & x<y \\ 
    \end{cases}.
\end{equation}
\end{widetext}
The Hamiltonian can be written in terms of eigenstates by $H(k_{j})=E \left| \chi_{c}(k_{i}) \right\rangle \left\langle \chi_{c}(k_{i})\right|-E \left| \chi_{c}(k_{i})\right\rangle \left\langle\chi_{c}(k_{i})\right |$, where $i=+$ for right-moving state and $i=-$ for left-moving state. Using this relation and completeness relation $1= \left| \chi_{c}(k_{i}) \right\rangle \left\langle \chi_{c}(k_{i})\right|+ \left| \chi_{c}(k_{i})\right\rangle \left\langle\chi_{c}(k_{i})\right |$, the real space Green's function can be written as. \par  
\begin{equation}
    G(x,y)=
     \begin{cases}
        -\frac{i L_{x} E}{v} \frac{\left|\chi_{c}(k_{+})\right \rangle \left \langle \chi_{c}(k_{+})\right|}{\sqrt{E^2-b_{e}^{2}}}e^{\frac{i \sqrt{E^2-b_{e}^2}(x-y)}{v}} & x>y \\
         -\frac{i L_{x} E}{v} \frac{\left|\chi_{c}(k_{-})\right \rangle \left \langle \chi_{c}(k_{-})\right|}{\sqrt{E^2-b_{e}^{2}}}e^{-\frac{i \sqrt{E^2-b_{e}^2}(x-y)}{v}} & x<y \\
    \end{cases}.
\end{equation}
Using the relation between incoming wave $\left|\chi_{c}(k_{+})\right \rangle$ with positive energy and momentum and final wave $\left|\chi_{out}(k)\right \rangle$, we can write $\left|\chi_{out}(k)\right \rangle=\left|\chi_{c}(k_{+})\right \rangle+GT\left|\chi_{c}(k_{+})\right \rangle$. The  reflection and transmission coefficients can be related to the T matrix and eigenstate for given energy. The reflection coefficient is \par 
\begin{equation}
    r=-\frac{i L_{x} E}{v \sqrt{E^2-b_{e}^{2}}} \left\langle\chi_{c}(k_{-})\right| T \left|\chi_{c}(k_{+})\right \rangle.
\end{equation}
We can simplify the reflection coefficient to \par 
\begin{equation}
    r=\frac{- 4 i \omega b_{e}}{(4-\omega^2)\sqrt{E^2-b^{2}_{e}}+4i \omega E}.
\end{equation}
Additionally, for single channel helical conductor, the transmission probability $T_{pot}$ is related to the reflection coefficient $r$ by $T_{pot}=|t|^{2}=1-|r|^{2}$. The transmission probability therefore is
\begin{equation}\label{eq:potentialtransmission}
    T_{pot}=\frac{E^2-b_{e}^2}{E^2-\left(\frac{4-\omega^2}{4+\omega^2}\right)^2b_{e}^2}.
\end{equation}
As expected, this transmission probability vanishes for $E=b_e$, \textit{i.e.} when the Zeeman gap on the edge ($2b_e$) is equal to the spin splitting at the Fermi energy ($2E$). For small edge Zeeman fields, it behaves as $T_{imp}\propto b_e^2/E^2$. This is the main lesson of this section: unlike the case of midgap states, potential disorder cannot lead to non-analytic behavior of the transmission coefficient at small fields. The reason for that is the existence of zero-field spin splitting for the edge states, which is absent for the midgap levels, as they are degenerate in the absence of time-reversal breaking due to the Kramers theorem. 

 \begin{figure}[htb]\label{fig:Tpot}
\begin{minipage}{8.6cm}
\begin{tikzpicture}
  \node (img)  {\includegraphics[width=7.6cm]{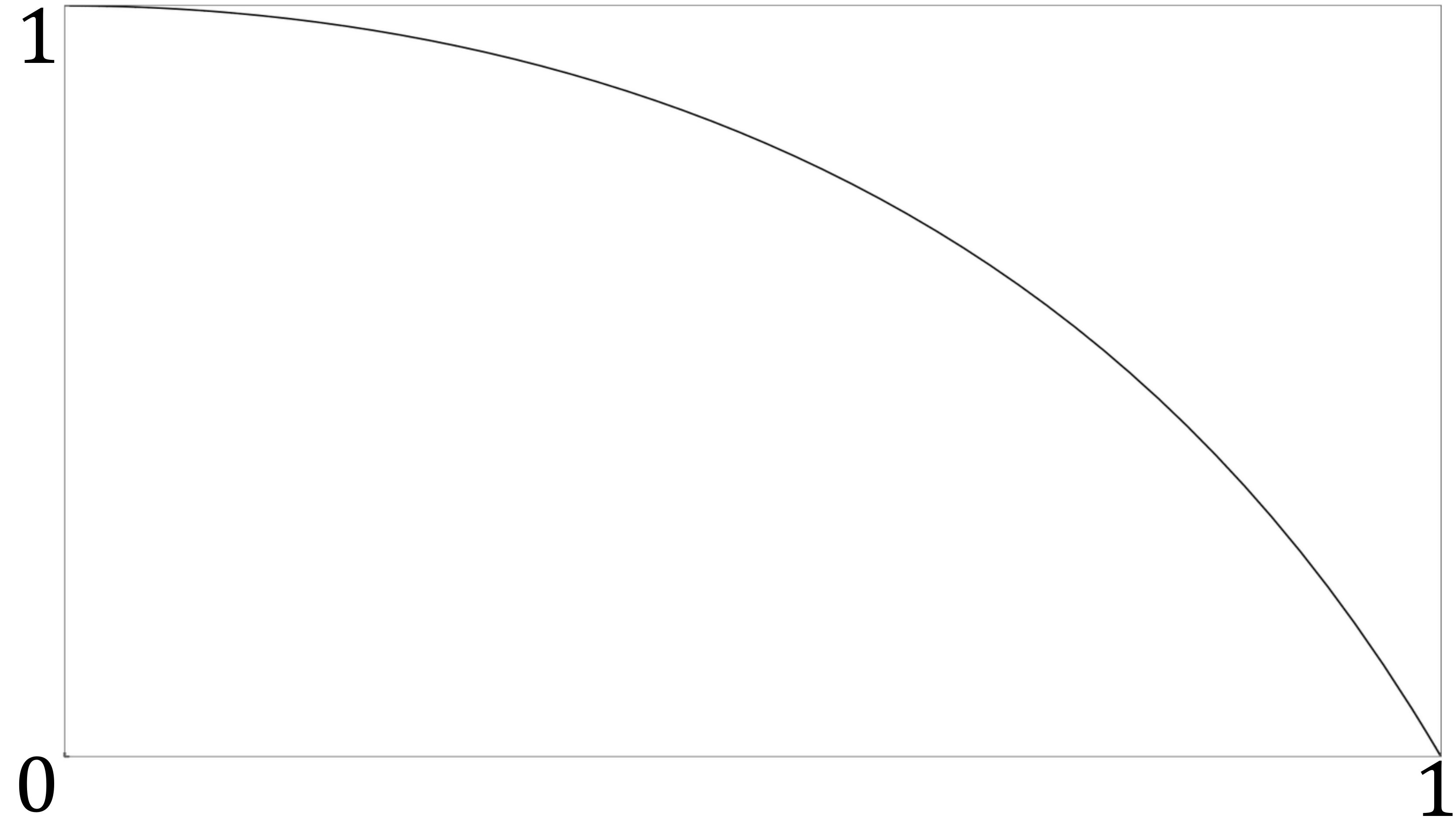}};
  \node[below=of img, node distance=0cm, yshift=1cm,font=\color{black}] { $b_{e}/E$ };
  \node[left=of img, node distance=0cm, rotate=90, anchor=center,yshift=-0.7cm,font=\color{black}] {$T_{pot}$};
 \end{tikzpicture}
\end{minipage}
\caption{\label{pot} The transmission probability as a function of magnetic field $b$. The dimensionless parameter $\omega$ is chosen to be $\omega=1$.}
\end{figure}
\clearpage
\bibliography{resimp}
\bibliographystyle{apsrev4-2}
\end{document}